\documentclass[prd,twocolumn,nofootinbib,reprint]{revtex4-1}
\usepackage{amsmath,amssymb,comment,url,hyperref,graphicx,color,fancybox}
\begin{document}
\rightline{DESY 19-122}
\author{Ryosuke Sato}
\title{Simple Gradient Flow Equation for the Bounce Solution}
\affiliation{\vspace{2mm} Deutsches Elektronen-Synchrotron (DESY), Notkestra\ss e 85, D-22607 Hamburg, Germany}
\date{\today}
\begin{abstract}
\vspace{1mm}
Motivated by the recent work of Chigusa, Moroi, and Shoji \cite{Chigusa:2019wxb},
we propose a new simple gradient flow equation to derive the bounce solution which contributes to the decay of the false vacuum.
Our discussion utilizes the discussion of Coleman, Glaser, and Martin \cite{Coleman:1977th}
and we solve a minimization problem of the kinetic energy while fixing the potential energy.
The bounce solution is derived as a scale-transformed of the solution of this problem.
We also show that the convergence of our method is robust against a choice of the initial configuration.
\end{abstract}
\maketitle

\section{Introduction}
The decay of the false vacua is an important topic in particle physics and cosmology.
The decay rate of the false vacua can be calculated from ``the imaginary part'' of the Euclidean path integral \cite{Coleman:1977py}\footnote{
For earlier discussions, see, \textit{e.g.,} Refs.~\cite{Lee:1974ma, Frampton:1976kf}.
}.
In the path integral formalism, we can see that the main contribution comes from the bounce solution $\phi_B$
which is a non-trivial solution of the equation of motion with the least action.
Thus, the bounce solution plays a crucial role in the decay of the false vacua.
To calculate the bounce solution, we have to solve the equation of motion with the boundary condition at infinity.
In general, it is not easy to calculate the bounce solution, and this is particularly the case for models with multi scalar fields.

Several algorithms to calculate the bounce action have been discussed so far,
\textit{e.g.},
gradient flow with modifications \cite{Claudson:1983et, Cline:1998rc, Cline:1999wi},
modified actions which have the bounce solution as a local minimum \cite{Kusenko:1995jv, Kusenko:1996jn, Moreno:1998bq, John:1998ip},
changing gradually a coefficient of the friction term (the second term on the LHS of Eq.~(\ref{eq:EOM2})) \cite{Konstandin:2006nd, Park:2010rh},
machine learning \cite{Jinno:2018dek, Piscopo:2019txs}
and so on.
Also, public codes to calculate the bounce solution are available, such as
\texttt{CosmoTransitions} \cite{Profumo:2010kp, Wainwright:2011kj},
\texttt{AnyBubble} \cite{Masoumi:2016wot},
and \texttt{BubbleProfiler} \cite{Akula:2016gpl, Athron:2019nbd}.
Some works discuss the bounce solution/action avoiding the direct calculation,
\textit{e.g.},
some approximations \cite{Johnson:2008kc, Masoumi:2017trx, Guada:2018jek},
upperbounds \cite{Dasgupta:1996qu, Sarid:1998sn, Brown:2017cca},
lowerbounds, \cite{Aravind:2014aza, Sato:2017iga, Brown:2017cca},
and an alternative formulation \cite{Espinosa:2018hue, Espinosa:2018voj, Espinosa:2018szu}.

One of the reasons for the technical difficulty is that the bounce solution is a saddle point of the action,
\textit{i.e.}, the bounce is not a stable solution of a simple minimization problem.
Recently, Chigusa, Shoji, and Moroi \cite{Chigusa:2019wxb} proposed a new method to obtain the bounce solution.
They proposed a gradient flow equation whose fixed point is the bounce solution.
Their flow equation has the gradient of the action and an additional term to lift up unstable direction around the bounce solution.
Motivated by Ref.~\cite{Chigusa:2019wxb}, in this paper, we propose a new simple flow equation.
Coleman, Glaser, and Martin (CGM) \cite{Coleman:1977th} showed that
the calculation of the bounce solution is equivalent to the minimization of the kinetic energy ${\cal T}$
while fixing the potential energy ${\cal V} < 0$.
This minimization problem can be naturally formulated in a flow equation.
In the end, the bounce solution is obtained as a scale-transformed of the solution of this problem.
In Sec.~\ref{sec:formulation}, we describe our formulation to calculate the bounce solution.
In Sec.~\ref{sec:example}, we discuss numerical analysis on several examples by using our flow equation,
and show that our flow equation works well.

\section{Formulation}\label{sec:formulation}
In this paper, we focus on the Euclidean action with $n$ scalar fields with the canonical kinetic term.
\begin{align}
{\cal S}[\phi] &= {\cal T}[\phi] + {\cal V}[\phi], \\
{\cal T}[\phi] &= \sum_{i=1}^n \int d^d x \frac{1}{2} (\nabla \phi_i)^2, \\
{\cal V}[\phi] &= \int d^d x V(\phi).
\end{align}
Here $d$ is the dimension of the space, and we assume $d$ is larger than 2.
The scalar potential $V$ satisfies $V(0) = 0$, $\partial V/\partial\phi_i = 0$,
all of the eigenvalues of the Hessian of $V$ at $\phi_i = 0$ are non-negative,
and $V$ is somewhere negative.

The bounce solution which contributes to the decay of the false vacuum
satisfies the equation of the motion and the boundary condition at infinity:
\begin{align}
-\nabla^2\phi_i + \frac{\partial V}{\partial \phi_i} &= 0, \label{eq:EOM}\\
\lim_{|x|\to\infty} \phi_i(x) &= 0. \label{eq:boundary}
\end{align}
Also, the bounce solution should be a non-trivial solution, \textit{i.e.}, $\exists i,x,~\phi_i(x) \neq 0$.
Thus,
\begin{align}
{\cal T}[\phi] > 0, \quad
{\cal V}[\phi] < 0. \label{eq:nontrivial}
\end{align}
Note that ${\cal V}[\phi] < 0$ is required in order for the bounce solution
to be an extremum under the scale transformation: $\phi_i(x) \to \phi_i(\lambda x)$.
See, \textit{e.g.}, Ref.~\cite{Coleman:1977th}.
The bounce solution has the least action among configurations which satisfy the above conditions Eqs.~(\ref{eq:EOM}, \ref{eq:boundary}, \ref{eq:nontrivial}).
It is known that the bounce solution has spherical symmetry \cite{Coleman:1977th, lopes1996radial, byeon2009symmetry, Blum:2016ipp}.
Therefore, Eq.~(\ref{eq:EOM}) can be simplified as
\begin{align}
-\frac{d^2\phi_i}{dr^2} - \frac{d-1}{r} \frac{d\phi_i}{dr} + \frac{\partial V}{\partial \phi_i} = 0. \label{eq:EOM2}
\end{align}

In order to discuss the bounce solution, CGM \cite{Coleman:1977th} introduced the reduced problem,
which is defined as
\textit{the problem of finding a configuration vanishing at infinity which minimizes $\cal T$ for some fixed negative $\cal V$.}
The existence of the solution of this problem is ensured by CGM's theorem B in Ref.~\cite{Coleman:1977th}
and Ref.~\cite{brezis1984minimum}.
Also, CGM's theorem A ensures that the bounce solution can be obtained as a scale-transformed of a solution of the reduced problem.
See the Appendix.
Here we solve the CGM's reduced problem by using a gradient flow equation.
We introduce functions $\varphi_i(r,\tau)$ and propose the following gradient flow equations:
\begin{align}
\frac{\partial}{\partial \tau} \varphi_i(r,\tau) &= \nabla^2 \varphi_i - \lambda[\phi] \frac{\partial V(\varphi)}{\partial \varphi_i} \label{eq:gradient flow equation}, \\
\lambda[\varphi] &= \frac{\displaystyle\sum_i \int_0^\infty dr r^{d-1} \frac{\partial V(\varphi)}{\partial \varphi_i} \nabla^2 \varphi_i}
{\displaystyle\sum_i \int_0^\infty dr r^{d-1} \left(\frac{\partial V(\varphi)}{\partial \varphi_i}\right)^2}.
\label{eq:lambda}
\end{align}
Here $\tau$ is ``the time'' for the flow of $\varphi$ and $\nabla^2 \varphi_i = \partial_r^2 \varphi_i + (d-1) (\partial_r \varphi)/r$.
We take the initial $\varphi(r,0)$ such that
\begin{align}
{\cal V}[\varphi] |_{\tau=0} < 0.
\end{align}
Note that $\lim_{r\to\infty}\varphi_i(r,\tau) = 0$ should hold in order for ${\cal V}[\phi]$ to be finite.
By using Eq.~(\ref{eq:gradient flow equation}) and Eq.~(\ref{eq:lambda}), we can show
\begin{align}
\frac{d}{d\tau}{\cal V}[\varphi] &= 0, \label{eq:dVds}\\
\frac{d}{d\tau}{\cal T}[\varphi] &\leq 0. \label{eq:dTds}
\end{align}
To show Eq.~(\ref{eq:dTds}), we used the following Cauchy-Schwarz inequality:
\begin{widetext}
\begin{align}
\left( \sum_i \int_0^\infty dr r^{d-1} (\nabla^2\varphi_i)^2\right)
\left( \sum_i \int_0^\infty dr r^{d-1} \left( \frac{\partial V(\varphi)}{\partial \varphi_i} \right)^2\right) 
~\geq~
\left( \sum_i \int_0^\infty dr r^{d-1} \frac{\partial V(\varphi)}{\partial \varphi_i} \nabla^2\varphi_i \right)^2. \label{eq:Cauchy Schwarz}
\end{align}
\end{widetext}
Also, we can see that the equalities of Eq.~(\ref{eq:dTds}) and Eq.~(\ref{eq:Cauchy Schwarz}) hold if and only if
\begin{align}
\nabla^2\varphi_i = \lambda \frac{\partial V(\varphi)}{\partial \varphi_i}
\end{align}
is satisfied.
Eqs.~(\ref{eq:dVds}, \ref{eq:dTds}) tell us that ${\cal T}[\varphi]$ monotonously decreases
while ${\cal V}[\varphi]$ is constant during the flow of $\varphi$.
In the limit of $\tau\to\infty$, $\varphi$ converges to a configuration
which satisfies $\nabla^2 \varphi_i - \lambda(\partial V(\varphi)/\partial\varphi_i) = 0$.
The convergence of $\varphi$ is guaranteed by the existence of the minimizer \cite{Coleman:1977th, brezis1984minimum}.
Note that this fixed point cannot be the false vacuum $\varphi_i = 0$
because ${\cal V}[\varphi]$ in the neighborhood of the false vacuum is positive 
and ${\cal V}[\varphi]$ is always negative during the flow.
As long as the initial condition is not fine-tuned,
$\varphi$ at $\tau\to\infty$ should be stable solution under the small perturbation,
\textit{i.e.}, ${\cal T}[\varphi]$ should be a local minimum under the small perturbation such that ${\cal V}[\phi]$ is not changed.
In principle, the reduced problem could have several local minima.
Physically, this case happens if there exist several directions of tunneling.
In this case, $\varphi$ at $\tau\to\infty$ depends on the initial condition,
and we can find the global minimum among those local minima.
The configuration which gives the smallest value of ${\cal T}$ is the solution of the CGM's reduced problem.

Let $\phi_i(r) (\equiv \lim_{\tau\to\infty} \varphi_i(r,\tau) )$ be the solution of the reduced problem,
and derive the bounce solution.
The bounce solution $\phi_B(r)$ can be obtained by a scale transformation of $\phi$ as
\begin{align}
\phi_B(r) = \phi(\lambda^{-1/2}r). \label{eq:bounce from gradient flow}
\end{align}
The above $\lambda$ is calculated as $\lim_{\tau\to\infty}\lambda[\varphi]$.
Although the CGM's theorem A ensures that this $\phi_B$ is the bounce solution,
let us see this more explicitly.
We can immediately see that (i) $\phi_B$ satisfies the EOM (Eq.~(\ref{eq:EOM}))
and (ii) $\lim_{r\to\infty} \phi_B(r) = 0$ because ${\cal V}[\phi_B]$ is finite.
Also, we can see that (iii) ${\cal S}$ has only one unstable direction around $\phi_B$.
Since $\phi_B$ is a scale-transformed of $\phi$,
$\phi_B$ is the global minimum of the action ${\cal S}$ if the potential energy ${\cal V}$ is fixed.
The direction in which ${\cal S}$ decreases is the direction which changes ${\cal V}[\phi]$, \textit{i.e.}, the scale transformation.
Therefore, $\phi_B$ which is defined in Eq.~(\ref{eq:bounce from gradient flow}) is the bounce solution.

An essential point of our method is that the negative eigenmode around the bounce solution can be related to the scale transformation.
By fixing the potential energy ${\cal V}$, we freeze fluctuation in this direction.
Note that a method which is proposed in Ref.~\cite{Claudson:1983et} also utilizes this property.

\section{Example}\label{sec:example}
In the previous section, we have seen that the CGM's reduced problem can be solved by the flow equation
Eq.~(\ref{eq:gradient flow equation}) and Eq.~(\ref{eq:lambda}),
and the bounce solution can be obtained from Eq.~(\ref{eq:bounce from gradient flow}).
In this section, we discuss numerical results for several examples, and show that our method works well.

\begin{figure}[h]
\includegraphics[width=0.9\hsize]{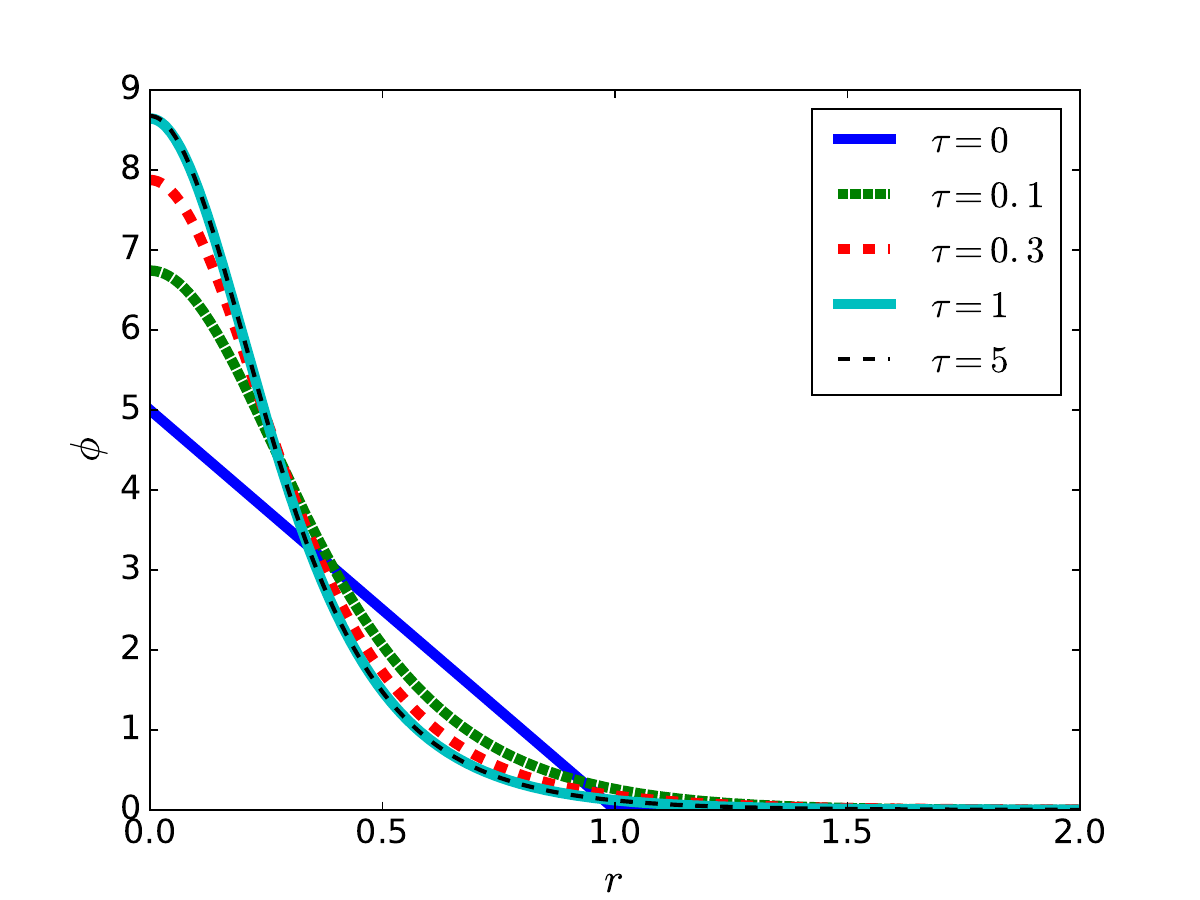}
\caption{
A flow of the field configuration with the potential Eq.~(\ref{eq:single potential}) with $d=4$
and the initial condition Eq.~(\ref{eq:single initial condition}).
 }\label{fig:singlefield evolution}
\end{figure}

\begin{figure}
\includegraphics[width=0.9\hsize]{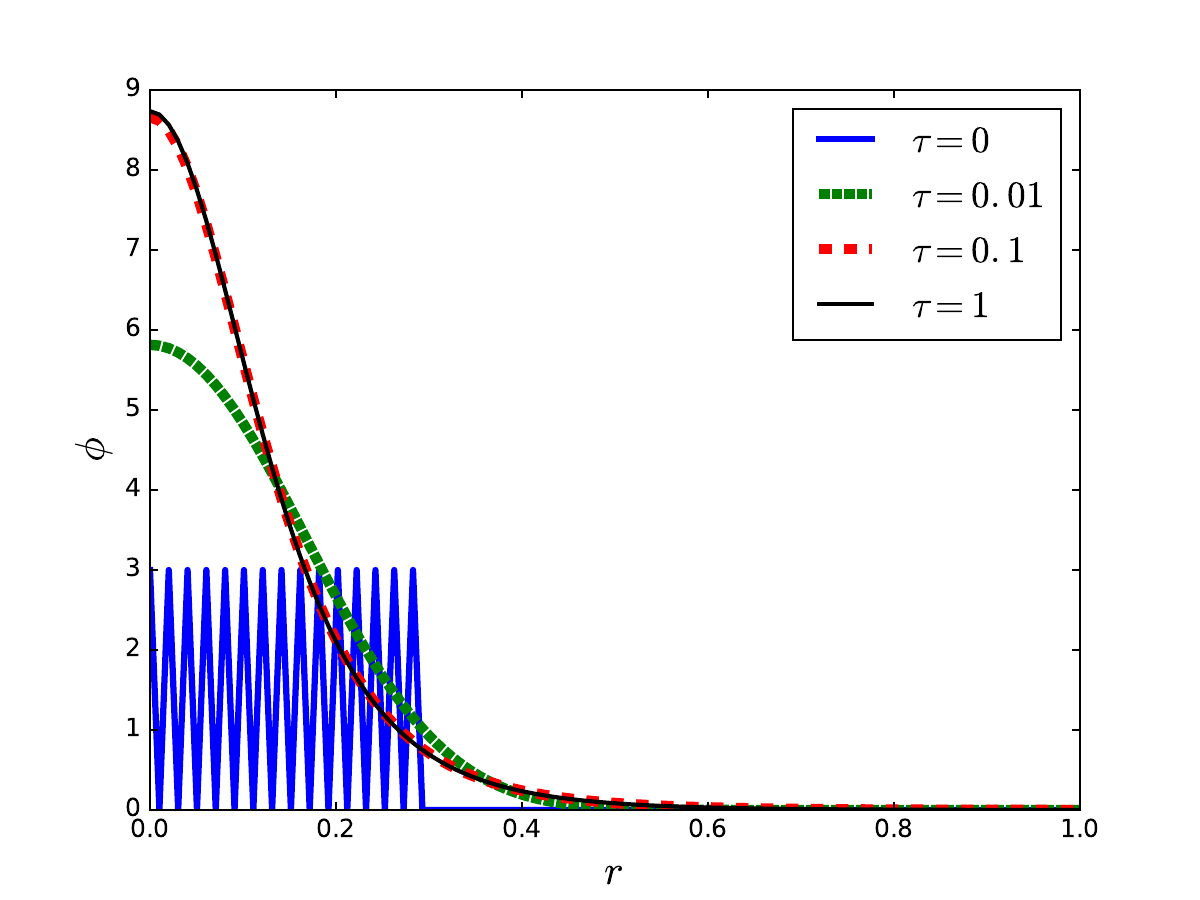}
\caption{
Same as Fig.~\ref{fig:singlefield evolution} except for the initial configuration.
}\label{fig:initial2}
\includegraphics[width=0.9\hsize]{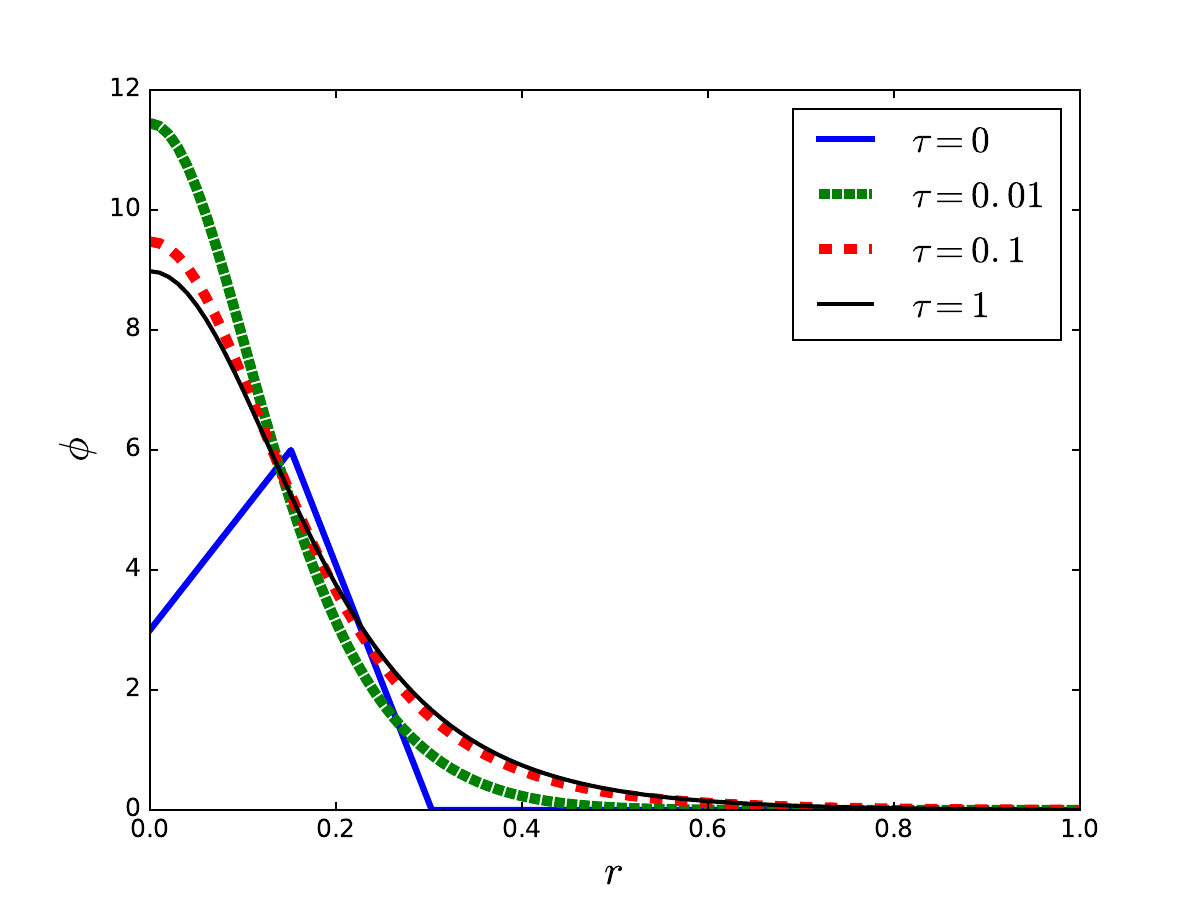}
\caption{
Same as Fig.~\ref{fig:singlefield evolution} except for the initial configuration.
}\label{fig:initial3}
\end{figure}

\begin{figure}[h]
\centering
\includegraphics[width=0.9\hsize]{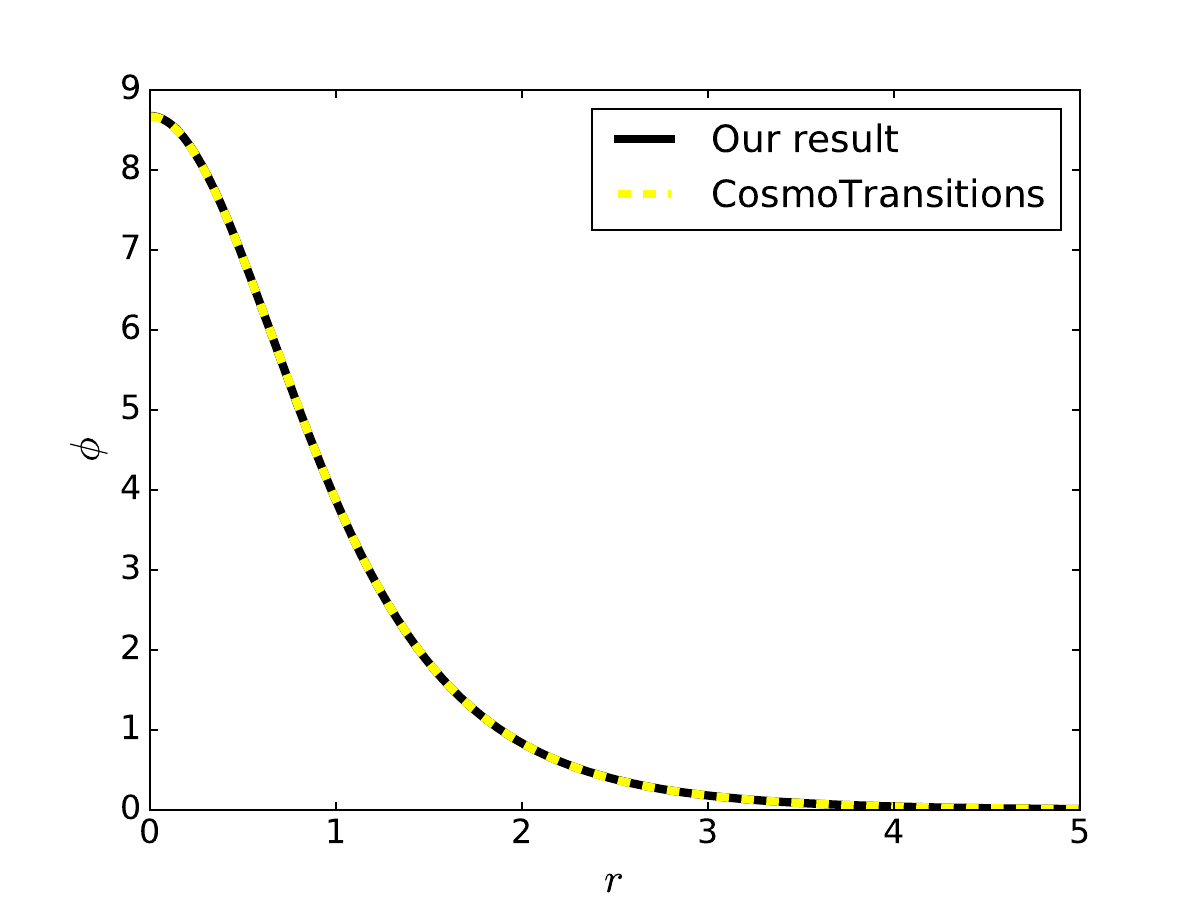}
\caption{
The black line is obtained from Eq.~(\ref{eq:bounce from gradient flow}) in the limit of large $\tau$.
The yellow dotted line is calculated by \texttt{CosmoTransitions}.
 }\label{fig:singlefield}
\end{figure}
First, let us take the following single scalar potential in $d=4$ Euclidean space.
\begin{align}
V(\phi) = \frac{1}{2} \phi^2 - \frac{1}{3}\phi^3. \label{eq:single potential}
\end{align}
We take the initial configuration at $\tau=0$ as
\begin{align}
\varphi(r,0) =
\begin{cases}
5(1-r) & (0\leq r \leq 1) \\
0 & (r>1)
\end{cases}. \label{eq:single initial condition}
\end{align}
The flow of this field configuration is shown in Fig.~\ref{fig:singlefield evolution}.
We can see the convergence of the configuration.
In Fig.~\ref{fig:initial2} and Fig.~\ref{fig:initial3}, we show the flow of the configuration from different initial conditions.
We can see that the convergence of the configuration is robust for different initial conditions.
The final configuration depends only on the value of ${\cal V}[\varphi]$,
and those different final results are connected with each other by an appropriate scale transformation.
By using this result, we can obtain the bounce solution from Eq.~(\ref{eq:bounce from gradient flow}).
We compare our bounce solution with the result by \texttt{CosmoTransitions} \cite{Wainwright:2011kj} in Fig.~\ref{fig:singlefield}.
We can see that the two results agree well and our method works.

\begin{figure}[h]
\includegraphics[width=0.9\hsize]{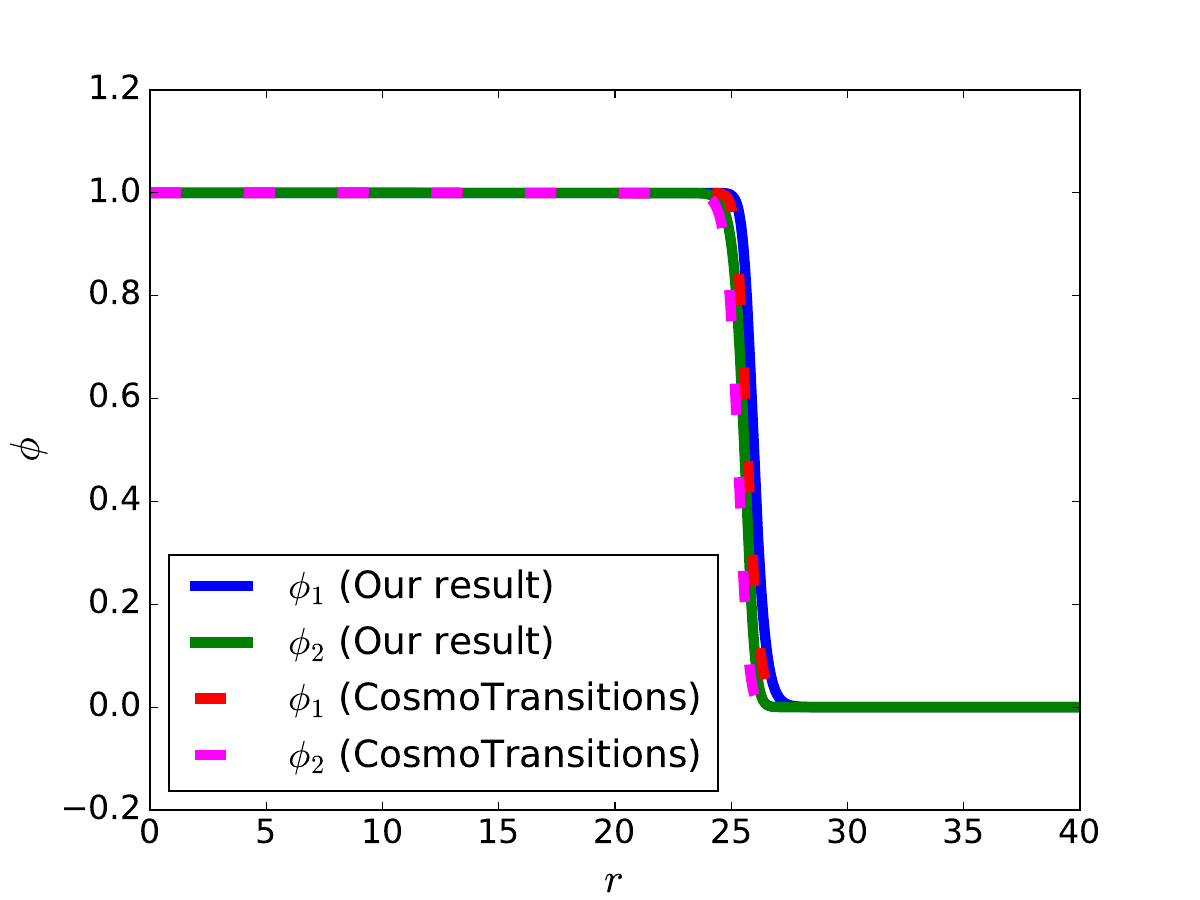}
\caption{
The bounce solution in the $r$-$\phi$ plane is shown by solid lines. The dashed lines are results of \texttt{CosmoTransitions}.
We take the potential Eq.~(\ref{eq:potential doublet scalar}) with $c=2$ in $d=4$ space.}\label{fig:doublescalar thickwall}
\end{figure}
\begin{figure}[h]
\includegraphics[width=0.9\hsize]{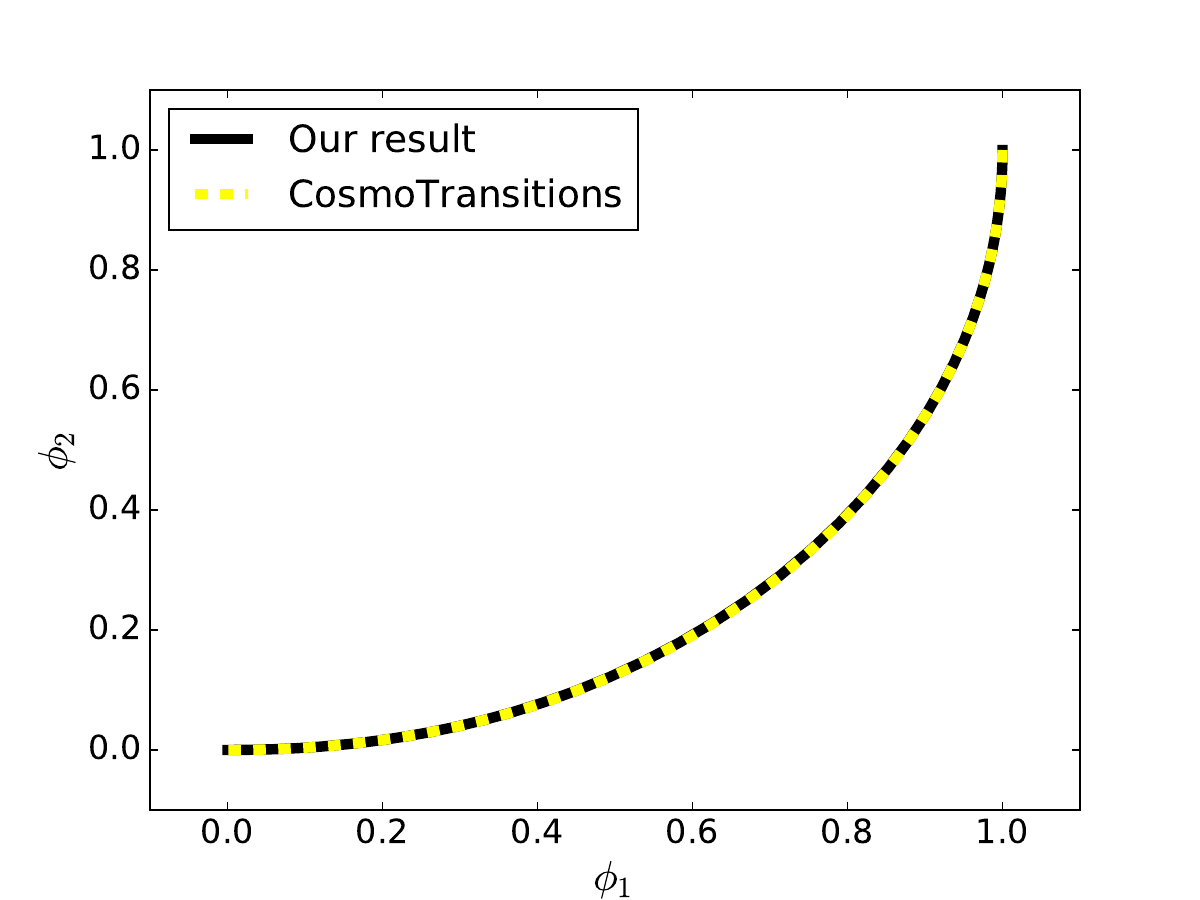}
\caption{The same bounce solution as Fig.~\ref{fig:doublescalar thickwall} in $\phi_1$-$\phi_2$ plane.}\label{fig:doublescalar thickwall 2}
\end{figure}
\begin{figure}[h]
\includegraphics[width=0.9\hsize]{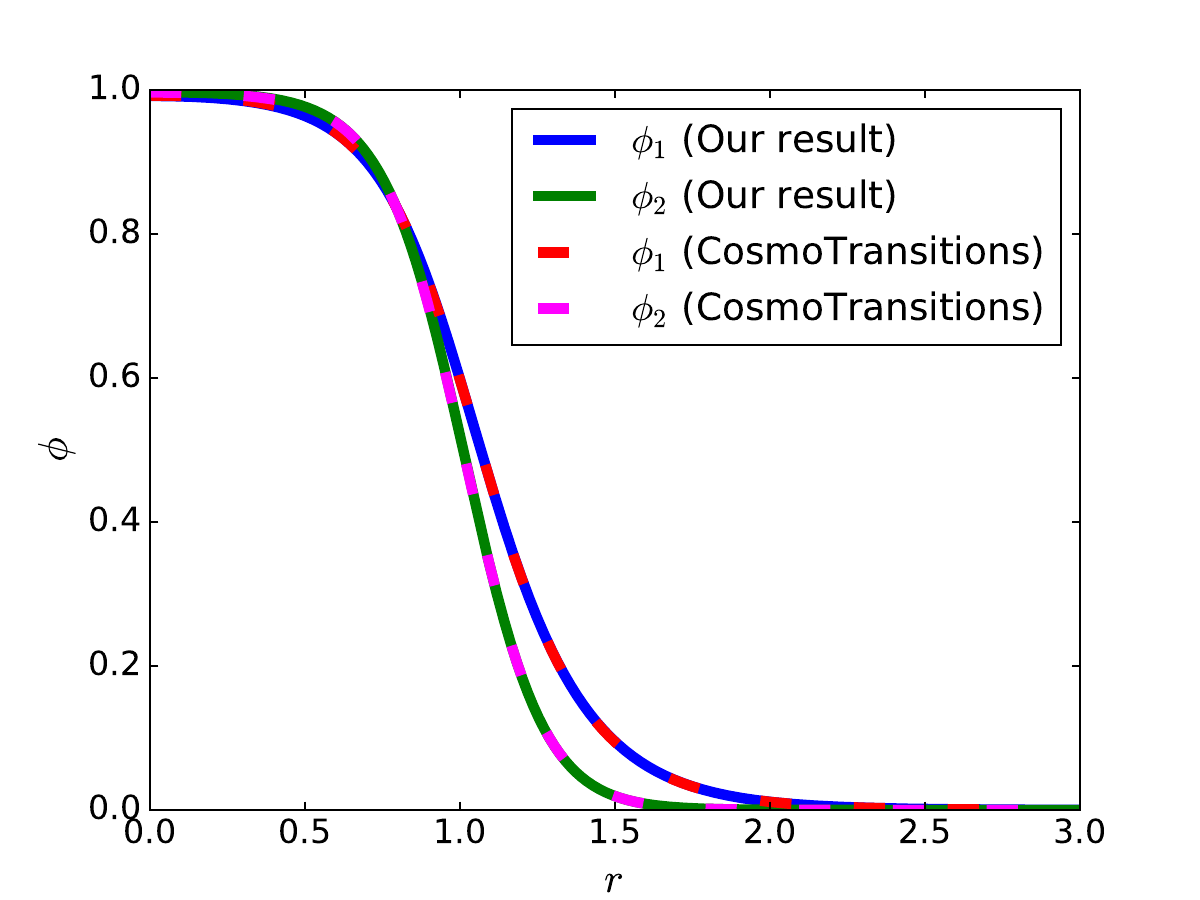}
\caption{
The bounce solution in the $r$-$\phi$ plane is shown by solid lines. The dashed lines are results of \texttt{CosmoTransitions}.
We take the potential Eq.~(\ref{eq:potential doublet scalar}) with $c=80$ in $d=4$ space.}\label{fig:doublescalar thinwall}
\end{figure}
\begin{figure}[h]
\includegraphics[width=0.9\hsize]{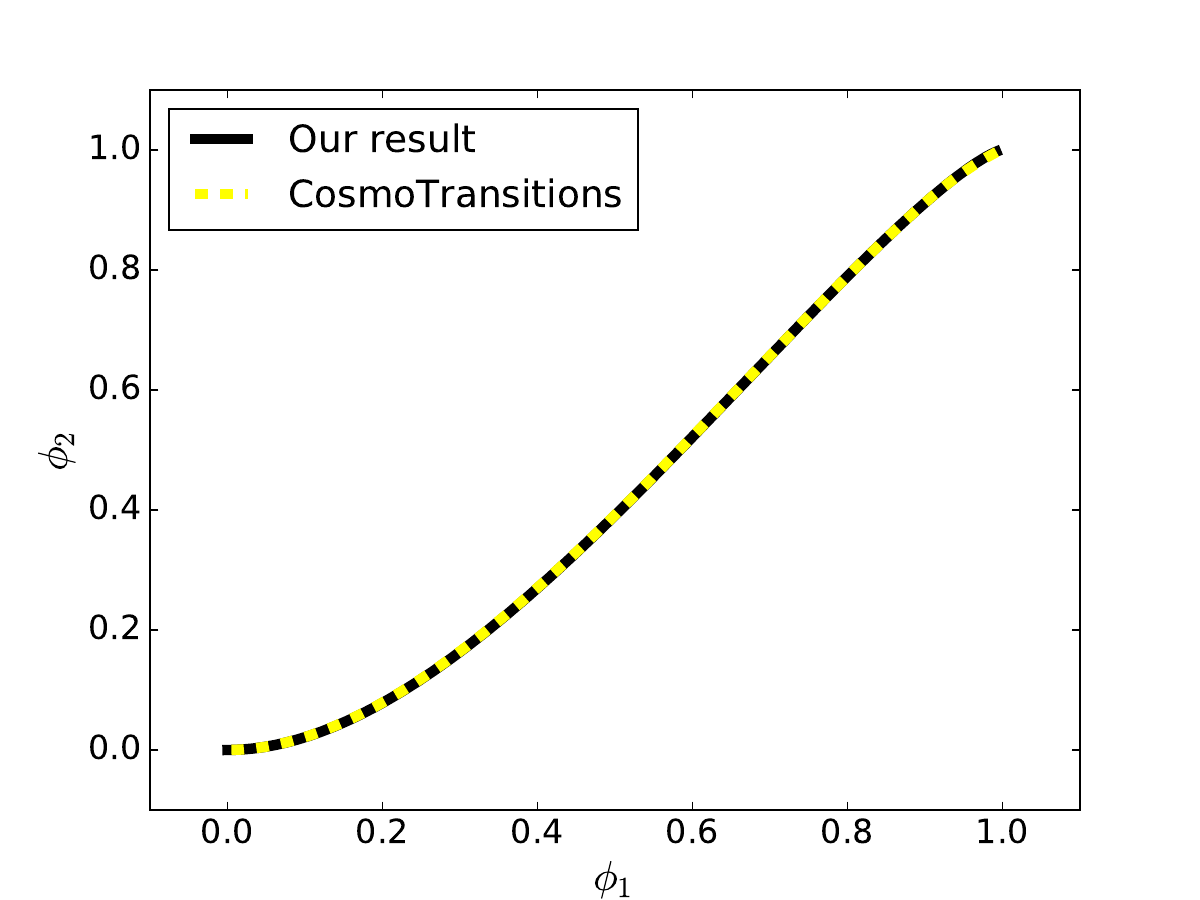}
\caption{The same bounce solution as Fig.~\ref{fig:doublescalar thinwall} in $\phi_1$-$\phi_2$ plane.}\label{fig:doublescalar thinwall 2}
\end{figure}
Next, let us discuss a case with two scalar fields.  We take the following potential:
\begin{align}
V =& (\phi_1^2 + 5 \phi_2^2) ( 5(\phi_1-1)^2 + (\phi_2-1)^2 ) \nonumber\\
   & + c\left( \frac{1}{4}\phi_2^4 - \frac{1}{3}\phi_2^3 \right). \label{eq:potential doublet scalar}
\end{align}
Again, we compare our bounce solutions with the results by \texttt{CosmoTransitions}.
The case with $c=2$ is shown in Figs.~\ref{fig:doublescalar thickwall} and \ref{fig:doublescalar thickwall 2},
and the case with $c=80$ in Figs.~\ref{fig:doublescalar thinwall} and \ref{fig:doublescalar thinwall 2}.
We can see that our result agrees with that of \texttt{CosmoTransitions}.

\section{Conclusion}
In this paper, motivated by a recent work of Chigusa, Shoji, and Moroi \cite{Chigusa:2019wxb},
we proposed a new simple gradient flow equation which is defined in Eq.~(\ref{eq:gradient flow equation}) and Eq.~(\ref{eq:lambda}).
Our flow equation solves the CGM's reduced problem \cite{Coleman:1977th}, \textit{i.e.},
the minimization problem of kinetic energy ${\cal T}$ while fixing potential energy ${\cal V}$.
This minimization problem can be naturally formulated in a flow equation,
and the bounce solution can be obtained as a scale-transformed of this solution as Eq.~(\ref{eq:bounce from gradient flow}).
Since our flow equation solves the minimization problem
and the existence of the minimizer is guaranteed by Refs.~\cite{Coleman:1977th, brezis1984minimum},
the convergence of this method is robust against the choice of initial configuration as long as ${\cal V}[\varphi] < 0$ is satisfied.
A numerical package using this method is presented in Ref.~\cite{Sato:2019wpo}\footnote{
See also \url{https://github.com/rsato64/SimpleBounce}.
}. 
This package calculates the Euclidean bounce action in ${\cal O}(0.1)$ s with ${\cal O}(0.1)~\%$ accuracy for models with 1--8 scalar field(s),
which is faster than \texttt{CosmoTransitions} \cite{Profumo:2010kp, Wainwright:2011kj},
\texttt{AnyBubble} \cite{Masoumi:2016wot},
and \texttt{BubbleProfiler} \cite{Akula:2016gpl, Athron:2019nbd}.

\section*{Acknowledgements}
The author thanks Takeo Moroi for useful discussions,
and also thanks Yu Hamada for pointing out a typo in Eq.~(\ref{eq:bounce from gradient flow}).

\appendix
\section{CGM's theorem A}\label{sec:theorem A}
In this Appendix, we briefly summarize the theorem A in Ref.~\cite{Coleman:1977th}.
We denote the solution of the reduced problem for given $\cal V$ as $\phi_{(\cal V)}$.
This theorem ensures that the bounce solution is given by a scale transformation of $\phi_{(\cal V)}$.

$\phi_{({\cal V}_0)}$ is a stationary point of ${\cal T}[\phi] + \lambda ({\cal V}[\phi]-{\cal V}_0)$,
where $\lambda$ is the Lagrange multiplier.
Thus, $\phi_{({\cal V}_0)}$ satisfies
\begin{align}
-\nabla^2 \phi_{({\cal V}_0)i} + \lambda \frac{\partial V}{\partial\phi_i} = 0. \label{eq:EOM reduced problem}
\end{align}
Here $\lambda$ should be appropriately chosen for the value of ${\cal V}_0$.
We define the following configuration $\phi_B$:
\begin{align}
\phi_B(x) = \phi_{({\cal V}_0)}(\lambda^{1/2} x). \label{eq:bounce scale-transformed}
\end{align}
We can see that this is the bounce solution.
First, by using Eqs.~(\ref{eq:EOM reduced problem}, \ref{eq:bounce scale-transformed}), we can check that $\phi_B$ satisfies the EOM Eq.~(\ref{eq:EOM}).
Next, let us show that the action of any non-trivial solution of Eq.~(\ref{eq:EOM}) is equal to or larger than ${\cal S}[\phi_B]$.
Let $\tilde\phi$ be a non-trivial solution of Eq.~(\ref{eq:EOM}).
The action of $\tilde\phi$ is extremized under the scale transformation of $\tilde\phi$. Therefore,
\begin{align}
(d-2) {\cal T}[\tilde\phi] + d {\cal V}[\tilde\phi] = 0. \label{eq:proof theoremA 1}
\end{align}
There exists a solution of the reduced problem for ${\cal V} = {\cal V}[\tilde\phi]$,
and the kinetic energy is not larger than ${\cal T}[\tilde\phi]$:
\begin{align}
{\cal T}[\phi_{({\cal V}[\tilde\phi])} ] \leq {\cal T}[\tilde\phi].\label{eq:proof theoremA 2}
\end{align}
${\cal T}[\phi_B]$ and ${\cal V}[\phi_B]$ are given as
\begin{align}
{\cal T}[\phi_B] &= \lambda^{1-d/2}  {\cal T}[\phi_{({\cal V}[\phi])} ], \label{eq:proof theoremA 3} \\
{\cal V}[\phi_B] &= \lambda^{-d/2} {\cal V}[\phi].
\end{align}
Here $\lambda \geq 1$ because of $(d-2) {\cal T}[\phi_B] + d {\cal V}[\phi_B] = 0$
and Eqs.~(\ref{eq:proof theoremA 1}, \ref{eq:proof theoremA 2}).
Thus, by using Eqs.~(\ref{eq:proof theoremA 2}, \ref{eq:proof theoremA 3}), we can show that 
\begin{align}
{\cal T}[\phi_B] \leq {\cal T}[\tilde\phi].
\end{align}
${\cal S} = (2/d){\cal T}$ is satisfied for solutions of Eq.~(\ref{eq:EOM}). Then,
\begin{align}
{\cal S}[\phi_B] \leq {\cal S}[\tilde\phi].
\end{align}
Thus, $\phi_B$ has the least action among the non-trivial solutions of the EOM.

\bibliography{ref}

\providecommand{\href}[2]{#2}\begingroup\raggedright\begin{thebibliography}{10}

\bibitem{Chigusa:2019wxb}
S.~Chigusa, T.~Moroi, and Y.~Shoji, ``{Bounce Configuration from Gradient
  Flow},'' \href{http://dx.doi.org/10.1016/j.physletb.2019.135115}{{\em Phys.
  Lett.} {\bfseries B800} (2020) 135115},
\href{http://arxiv.org/abs/1906.10829}{{\ttfamily arXiv:1906.10829 [hep-ph]}}.

\bibitem{Coleman:1977th}
S.~R. Coleman, V.~Glaser, and A.~Martin, ``{Action Minima Among Solutions to a
  Class of Euclidean Scalar Field Equations},''
\href{http://dx.doi.org/10.1007/BF01609421}{{\em Commun. Math. Phys.}
  {\bfseries 58} (1978) 211--221}.

\bibitem{Coleman:1977py}
S.~R. Coleman, ``{The Fate of the False Vacuum. 1. Semiclassical Theory},''
  \href{http://dx.doi.org/10.1103/PhysRevD.15.2929,
  10.1103/PhysRevD.16.1248}{{\em Phys. Rev.} {\bfseries D15} (1977)
  2929--2936}.
[Erratum: Phys. Rev.D16,1248(1977)].

\bibitem{Lee:1974ma}
T.~D. Lee and G.~C. Wick, ``{Vacuum Stability and Vacuum Excitation in a Spin 0
  Field Theory},''
\href{http://dx.doi.org/10.1103/PhysRevD.9.2291}{{\em Phys. Rev.} {\bfseries
  D9} (1974) 2291--2316}.

\bibitem{Frampton:1976kf}
P.~H. Frampton, ``{Vacuum Instability and Higgs Scalar Mass},''
  \href{http://dx.doi.org/10.1103/PhysRevLett.37.1378,
  10.1103/PhysRevLett.37.1716.2}{{\em Phys. Rev. Lett.} {\bfseries 37} (1976)
  1378}.
[Erratum: Phys. Rev. Lett.37,1716(1976)].

\bibitem{Claudson:1983et}
M.~Claudson, L.~J. Hall, and I.~Hinchliffe, ``{Low-Energy Supergravity: False
  Vacua and Vacuous Predictions},''
\href{http://dx.doi.org/10.1016/0550-3213(83)90556-4}{{\em Nucl. Phys.}
  {\bfseries B228} (1983) 501--528}.

\bibitem{Cline:1998rc}
J.~M. Cline, J.~R. Espinosa, G.~D. Moore, and A.~Riotto, ``{String mediated
  electroweak baryogenesis: A Critical analysis},''
  \href{http://dx.doi.org/10.1103/PhysRevD.59.065014}{{\em Phys. Rev.}
  {\bfseries D59} (1999) 065014},
\href{http://arxiv.org/abs/hep-ph/9810261}{{\ttfamily arXiv:hep-ph/9810261
  [hep-ph]}}.

\bibitem{Cline:1999wi}
J.~M. Cline, G.~D. Moore, and G.~Servant, ``{Was the electroweak phase
  transition preceded by a color broken phase?},''
  \href{http://dx.doi.org/10.1103/PhysRevD.60.105035}{{\em Phys. Rev.}
  {\bfseries D60} (1999) 105035},
\href{http://arxiv.org/abs/hep-ph/9902220}{{\ttfamily arXiv:hep-ph/9902220
  [hep-ph]}}.

\bibitem{Kusenko:1995jv}
A.~Kusenko, ``{Improved action method for analyzing tunneling in quantum field
  theory},'' \href{http://dx.doi.org/10.1016/0370-2693(95)00994-V}{{\em Phys.
  Lett.} {\bfseries B358} (1995) 51--55},
\href{http://arxiv.org/abs/hep-ph/9504418}{{\ttfamily arXiv:hep-ph/9504418
  [hep-ph]}}.

\bibitem{Kusenko:1996jn}
A.~Kusenko, P.~Langacker, and G.~Segre, ``{Phase transitions and vacuum
  tunneling into charge and color breaking minima in the MSSM},''
  \href{http://dx.doi.org/10.1103/PhysRevD.54.5824}{{\em Phys. Rev.} {\bfseries
  D54} (1996) 5824--5834},
\href{http://arxiv.org/abs/hep-ph/9602414}{{\ttfamily arXiv:hep-ph/9602414
  [hep-ph]}}.

\bibitem{Moreno:1998bq}
J.~M. Moreno, M.~Quiros, and M.~Seco, ``{Bubbles in the supersymmetric standard
  model},'' \href{http://dx.doi.org/10.1016/S0550-3213(98)00283-1}{{\em Nucl.
  Phys.} {\bfseries B526} (1998) 489--500},
\href{http://arxiv.org/abs/hep-ph/9801272}{{\ttfamily arXiv:hep-ph/9801272
  [hep-ph]}}.

\bibitem{John:1998ip}
P.~John, ``{Bubble wall profiles with more than one scalar field: A Numerical
  approach},'' \href{http://dx.doi.org/10.1016/S0370-2693(99)00272-5}{{\em
  Phys. Lett.} {\bfseries B452} (1999) 221--226},
\href{http://arxiv.org/abs/hep-ph/9810499}{{\ttfamily arXiv:hep-ph/9810499
  [hep-ph]}}.

\bibitem{Konstandin:2006nd}
T.~Konstandin and S.~J. Huber, ``{Numerical approach to multi dimensional phase
  transitions},'' \href{http://dx.doi.org/10.1088/1475-7516/2006/06/021}{{\em
  JCAP} {\bfseries 0606} (2006) 021},
\href{http://arxiv.org/abs/hep-ph/0603081}{{\ttfamily arXiv:hep-ph/0603081
  [hep-ph]}}.

\bibitem{Park:2010rh}
J.-h. Park, ``{Constrained potential method for false vacuum decays},''
  \href{http://dx.doi.org/10.1088/1475-7516/2011/02/023}{{\em JCAP} {\bfseries
  1102} (2011) 023},
\href{http://arxiv.org/abs/1011.4936}{{\ttfamily arXiv:1011.4936 [hep-ph]}}.

\bibitem{Jinno:2018dek}
R.~Jinno, ``{Machine learning for bounce calculation},''
\href{http://arxiv.org/abs/1805.12153}{{\ttfamily arXiv:1805.12153 [hep-th]}}.

\bibitem{Piscopo:2019txs}
M.~L. Piscopo, M.~Spannowsky, and P.~Waite, ``{Solving differential equations
  with neural networks: Applications to the calculation of cosmological phase
  transitions},'' \href{http://dx.doi.org/10.1103/PhysRevD.100.016002}{{\em
  Phys. Rev.} {\bfseries D100} no.~1, (2019) 016002},
\href{http://arxiv.org/abs/1902.05563}{{\ttfamily arXiv:1902.05563 [hep-ph]}}.

\bibitem{Profumo:2010kp}
S.~Profumo, L.~Ubaldi, and C.~Wainwright, ``{Singlet Scalar Dark Matter:
  monochromatic gamma rays and metastable vacua},''
  \href{http://dx.doi.org/10.1103/PhysRevD.82.123514}{{\em Phys. Rev.}
  {\bfseries D82} (2010) 123514},
\href{http://arxiv.org/abs/1009.5377}{{\ttfamily arXiv:1009.5377 [hep-ph]}}.

\bibitem{Wainwright:2011kj}
C.~L. Wainwright, ``{CosmoTransitions: Computing Cosmological Phase Transition
  Temperatures and Bubble Profiles with Multiple Fields},''
  \href{http://dx.doi.org/10.1016/j.cpc.2012.04.004}{{\em Comput. Phys.
  Commun.} {\bfseries 183} (2012) 2006--2013},
\href{http://arxiv.org/abs/1109.4189}{{\ttfamily arXiv:1109.4189 [hep-ph]}}.

\bibitem{Masoumi:2016wot}
A.~Masoumi, K.~D. Olum, and B.~Shlaer, ``{Efficient numerical solution to
  vacuum decay with many fields},''
  \href{http://dx.doi.org/10.1088/1475-7516/2017/01/051}{{\em JCAP} {\bfseries
  1701} no.~01, (2017) 051},
\href{http://arxiv.org/abs/1610.06594}{{\ttfamily arXiv:1610.06594 [gr-qc]}}.

\bibitem{Akula:2016gpl}
S.~Akula, C.~Bal\'{a}zs, and G.~A. White, ``{Semi-analytic techniques for
  calculating bubble wall profiles},''
  \href{http://dx.doi.org/10.1140/epjc/s10052-016-4519-5}{{\em Eur. Phys. J.}
  {\bfseries C76} no.~12, (2016) 681},
\href{http://arxiv.org/abs/1608.00008}{{\ttfamily arXiv:1608.00008 [hep-ph]}}.

\bibitem{Athron:2019nbd}
P.~Athron, C.~Bal\'{a}zs, M.~Bardsley, A.~Fowlie, D.~Harries, and G.~White,
  ``{BubbleProfiler: finding the field profile and action for cosmological
  phase transitions},'' \href{http://dx.doi.org/10.1016/j.cpc.2019.05.017}{{\em
  Comput. Phys. Commun.} {\bfseries 244} (2019) 448--468},
\href{http://arxiv.org/abs/1901.03714}{{\ttfamily arXiv:1901.03714 [hep-ph]}}.

\bibitem{Johnson:2008kc}
M.~C. Johnson and M.~Larfors, ``{Field dynamics and tunneling in a flux
  landscape},'' \href{http://dx.doi.org/10.1103/PhysRevD.78.083534}{{\em Phys.
  Rev.} {\bfseries D78} (2008) 083534},
\href{http://arxiv.org/abs/0805.3705}{{\ttfamily arXiv:0805.3705 [hep-th]}}.

\bibitem{Masoumi:2017trx}
A.~Masoumi, K.~D. Olum, and J.~M. Wachter, ``{Approximating tunneling rates in
  multi-dimensional field spaces},''
  \href{http://dx.doi.org/10.1088/1475-7516/2017/10/022}{{\em JCAP} {\bfseries
  1710} no.~10, (2017) 022},
\href{http://arxiv.org/abs/1702.00356}{{\ttfamily arXiv:1702.00356 [gr-qc]}}.

\bibitem{Guada:2018jek}
V.~Guada, A.~Maiezza, and M.~Nemev\v{s}ek, ``{Multifield Polygonal Bounces},''
  \href{http://dx.doi.org/10.1103/PhysRevD.99.056020}{{\em Phys. Rev.}
  {\bfseries D99} no.~5, (2019) 056020},
\href{http://arxiv.org/abs/1803.02227}{{\ttfamily arXiv:1803.02227 [hep-th]}}.

\bibitem{Dasgupta:1996qu}
I.~Dasgupta, ``{Estimating vacuum tunneling rates},''
  \href{http://dx.doi.org/10.1016/S0370-2693(96)01685-1}{{\em Phys. Lett.}
  {\bfseries B394} (1997) 116--122},
\href{http://arxiv.org/abs/hep-ph/9610403}{{\ttfamily arXiv:hep-ph/9610403
  [hep-ph]}}.

\bibitem{Sarid:1998sn}
U.~Sarid, ``{Tools for tunneling},''
  \href{http://dx.doi.org/10.1103/PhysRevD.58.085017}{{\em Phys. Rev.}
  {\bfseries D58} (1998) 085017},
\href{http://arxiv.org/abs/hep-ph/9804308}{{\ttfamily arXiv:hep-ph/9804308
  [hep-ph]}}.

\bibitem{Brown:2017cca}
A.~R. Brown, ``{Thin-wall approximation in vacuum decay: A lemma},''
  \href{http://dx.doi.org/10.1103/PhysRevD.97.105002}{{\em Phys. Rev.}
  {\bfseries D97} no.~10, (2018) 105002},
\href{http://arxiv.org/abs/1711.07712}{{\ttfamily arXiv:1711.07712 [hep-th]}}.

\bibitem{Aravind:2014aza}
A.~Aravind, D.~Lorshbough, and S.~Paban, ``{Lower bound for the multifield
  bounce action},'' \href{http://dx.doi.org/10.1103/PhysRevD.89.103535}{{\em
  Phys. Rev.} {\bfseries D89} no.~10, (2014) 103535},
\href{http://arxiv.org/abs/1401.1230}{{\ttfamily arXiv:1401.1230 [hep-th]}}.

\bibitem{Sato:2017iga}
R.~Sato and M.~Takimoto, ``{Absolute Lower Bound on the Bounce Action},''
  \href{http://dx.doi.org/10.1103/PhysRevLett.120.091802}{{\em Phys. Rev.
  Lett.} {\bfseries 120} no.~9, (2018) 091802},
\href{http://arxiv.org/abs/1707.01099}{{\ttfamily arXiv:1707.01099 [hep-ph]}}.

\bibitem{Espinosa:2018hue}
J.~R. Espinosa, ``{A Fresh Look at the Calculation of Tunneling Actions},''
  \href{http://dx.doi.org/10.1088/1475-7516/2018/07/036}{{\em JCAP} {\bfseries
  1807} no.~07, (2018) 036},
\href{http://arxiv.org/abs/1805.03680}{{\ttfamily arXiv:1805.03680 [hep-th]}}.

\bibitem{Espinosa:2018voj}
J.~R. Espinosa, ``{Fresh look at the calculation of tunneling actions including
  gravitational effects},''
  \href{http://dx.doi.org/10.1103/PhysRevD.100.104007}{{\em Phys. Rev.}
  {\bfseries D100} no.~10, (2019) 104007},
\href{http://arxiv.org/abs/1808.00420}{{\ttfamily arXiv:1808.00420 [hep-th]}}.

\bibitem{Espinosa:2018szu}
J.~R. Espinosa and T.~Konstandin, ``{A Fresh Look at the Calculation of
  Tunneling Actions in Multi-Field Potentials},''
  \href{http://dx.doi.org/10.1088/1475-7516/2019/01/051}{{\em JCAP} {\bfseries
  1901} no.~01, (2019) 051},
\href{http://arxiv.org/abs/1811.09185}{{\ttfamily arXiv:1811.09185 [hep-th]}}.

\bibitem{lopes1996radial}
O.~Lopes, ``Radial symmetry of minimizers for some translation and rotation
  invariant functionals,'' \href{http://dx.doi.org/10.1006/jdeq.1996.0015}{{\em
  Journal of differential equations} {\bfseries 124} no.~2, (1996) 378--388}.

\bibitem{byeon2009symmetry}
J.~Byeon, L.~Jeanjean, and M.~Mari{\c{s}}, ``Symmetry and monotonicity of least
  energy solutions,'' \href{http://dx.doi.org/10.1007/s00526-009-0238-1}{{\em
  Calculus of Variations and Partial Differential Equations} {\bfseries 36}
  no.~4, (2009) 481--492}, \href{http://arxiv.org/abs/0806.0299}{{\ttfamily
  arXiv:0806.0299 [math.AP]}}.

\bibitem{Blum:2016ipp}
K.~Blum, M.~Honda, R.~Sato, M.~Takimoto, and K.~Tobioka, ``{O($N$) Invariance
  of the Multi-Field Bounce},''
  \href{http://dx.doi.org/10.1007/JHEP05(2017)109,
  10.1007/JHEP06(2017)060}{{\em JHEP} {\bfseries 05} (2017) 109},
  \href{http://arxiv.org/abs/1611.04570}{{\ttfamily arXiv:1611.04570
  [hep-th]}}.
[Erratum: JHEP06,060(2017)].

\bibitem{brezis1984minimum}
H.~Brezis and E.~H. Lieb, ``Minimum action solutions of some vector field
  equations,'' \href{http://dx.doi.org/10.1007/BF01217349}{{\em Commun. Math.
  Phys.} {\bfseries 96} no.~1, (1984) 97--113}.
  \url{http://projecteuclid.org/euclid.cmp/1103941720}.

\bibitem{Sato:2019wpo}
R.~Sato, ``{SimpleBounce : a simple package for the false vacuum decay},''
\href{http://arxiv.org/abs/1908.10868}{{\ttfamily arXiv:1908.10868 [hep-ph]}}.

\end{thebibliography}\endgroup
\bibliographystyle{utphys}
\end{document}